\documentclass[%
 reprint,
 prc,
 amsmath,amssymb,
 aps,
 lengthcheck,
floatfix,
]{revtex4}

\usepackage{graphicx}
\usepackage{dcolumn}
\usepackage{bm}
\usepackage{hyperref}
\usepackage[utf8]{inputenc}
\usepackage{multirow}
\usepackage{soul}
\usepackage{float}
\usepackage{graphicx}
\usepackage{times}
\usepackage[normalem]{ulem}
\usepackage{color}
\usepackage{cellspace}
\usepackage{calrsfs}
\newcommand{\be}{\begin{equation}}
\newcommand{\ee}{\end{equation}}
\newcommand{\bea}{\begin{eqnarray}}
\newcommand{\eea}{\end{eqnarray}}

\newcommand{\comment}[1]{}
\renewcommand\sout{\bgroup \color{red} \ULdepth=-.5ex \ULset}
\def\simge{\mathrel{\rlap{\raise 0.511ex
     \hbox{$>$}}{\lower 0.511ex \hbox{$\sim$}}}}
\def\simle{\mathrel{\rlap{\raise 0.511ex
      \hbox{$<$}}{\lower 0.511ex \hbox{$\sim$}}}}

\begin{document}


\title{Nucleonic metamodelling in light of multimessenger, PREX-II and CREX data}

%
\author{C. Mondal$^1$}
\email{mondal@lpccaen.in2p3.fr}
\author{F. Gulminelli$^2$}
\email{gulminelli@lpccaen.in2p3.fr}

\affiliation{$^1$Laboratoire de Physique Corpusculaire, CNRS, ENSICAEN, UMR6534, Université de Caen Normandie,
F-14000, Caen Cedex, France}

\date{\today}

\begin{abstract} 
	The need of reconciling our understanding of the behavior of hadronic matter 
	across a wide range of densities, especially at the time when data from 
	multimessenger observations and novel experimental facilities 
	are flooding in, has provided new challenges to the nuclear models. 
	Particularly, the density dependence of the isovector channel of the nuclear energy functionals 
	seems hard to pin down if experiments like PREX-II (or PREX) and CREX are required 
	to be taken on the same footing. We put to test this anomaly in a 
	semi-agnostic modelling technique, by performing a full Bayesian analysis  
	of  static properties of neutron stars, together with global properties of 
	nuclei as binding energy, charge radii and neutron skin calculated at the semi-classical 
	level. Our results show that the interplay between bulk and surface properties, 
	and the importance of high order empirical parameters that effectively decouple 
	the subsaturation and the supersaturation density regime, might partially 
	explain the tension between the different measurements and observations.
	If the surface behaviors, however, are decoupled from the bulk properties, 
	we found a rather harmonious situation among experimental and observational 
	data. 
\end{abstract}


\maketitle

\section{Introduction}
The electroweak probe of the isovector channel of the nuclear
interaction obtained by studying the parity violating asymmetry in the elastic 
scattering channel, with the use of longitudinally polarized electrons 
as projectiles on neutron rich target nuclei \textit{e.g.} $^{208}$Pb 
(PREX, PREX-II) \cite{Abrahamya12,Adhikari21} or $^{48}$Ca (CREX), 
\cite{Adhikari22} has produced some 
very exciting discussions in recent times. Since the first run with 
$^{208}$Pb nucleus in Jefferson lab \cite{Abrahamya12}, many theoretical studies 
were conducted using the data as a constraining probe for model building. 
Its lack of precision, however, failed to induce any significant improvement 
in the modeling of the isovector sector of the nuclear interaction. The second run, 
referred as PREX-II, was able to reduce the uncertainty in the measurement of 
amplitude of parity violation quite significantly. Its direct inference 
on the neutron skin (connected directly to the weak charge distribution inside 
nuclei) is however in a rift with inferences made by alternative hadronic 
probes \cite{Krasznahorkay04, Klos07, Friedman09, Zenihiro10}. 
Furthermore, explaining results on dipole polarizability, amplitudes of parity 
violation in both CREX and PREX-II within the density functional theory was 
found to be in a spot of bother, and the anomaly with respect to our previous 
understanding of the density dependence of symmetry energy became even 
more prominent \cite{Reinhard21,Reinhard22,Yuksel22}. 
It was pointed out \cite{Typel10, Typel14,Tanaka21} that the standard understanding 
of the neutron skin through density functional theory 
and its connection to the density dependence of the symmetry energy, particularly the 
slope parameter ($L_{sym}$), might miss some beyond mean-field contribution.
However, the mean-field formalism has been extremely successful over the years in 
explaining a plethora of experimental data, and within this formalism a 
connection between the skin and the symmetry energy 
clearly exists \cite{Centelles09, Agrawal12, Pastore20, Mondal22b}. In addition to nuclear structure observables, 
heavy ion collision experiments have provided convincing constraints to the symmetry energy\cite{Tsang04, 
Danielewicz02, Adamczewski-Musch20}. 
Even more stringent constraints come from neutron star observational data, 
pouring in  during the past decade  \cite{Demorest10, Antoniadis13, Riley19, Miller19, 
Riley21, Miller21, Abbott17, Abbott18, Abbott19, LIGO15, 
Acernese14}. Those data  opened up 
new frontiers in the nuclear theory providing a formidable boost to the 
understanding of dense nuclear matter, which is typically beyond our reach 
in the terrestrial experimental facilities. 

This heavy supply of different types of data should help to pin 
down quite precisely the nuclear equation of state (EoS), 
and hopefully to connect it unequivocally to the underlying nuclear theory. 
Especially for this reason Bayesian studies have been employed 
quite frequently in recent times to extract the dense matter 
behavior in an agnostic way \cite{Steiner13, Margueron18a, Zhang18, 
Lim19, Traversi20, Biswas21a, Dinh-Thi21, 
Essick21, Somasundaram21, Ghosh22, Imam22, Mondal22a, Malik22, 
Volkel22, Huth22}. A certain degree of agnosticity 
is clearly required in order to extract the EoS from the astrophysical 
data in a model independent way. At the same time, the underlying correlations imparted on the 
theory by laboratory data at low densities coming from finite nuclei are 
also to be respected \cite{Roca-Maza18}. In the absence of a true theory that can 
explain the strong interaction from bottom up in the whole density 
regime covered by the experiments and the observations, functionals 
based on relativistic \cite{Walecka86} or non-relativistic zero-range \cite{Skyrme56,Skyrme59} 
and finite-range \cite{Decharge80} mean-field theory, and their multitude of extended 
versions are usually employed to build the nuclear EoS, 
see ref.\cite{Ryssens21}, and references therein, for recent developments. 
These complex modelings often suffer from shortage in flexibility, though 
some efforts are being made presently to employ them in Bayesian studies 
for finite nuclear properties\cite{Xu22}. To address this limitation, a more general 
meta-modeling technique based on a density expansion in terms of empirical 
parameters of infinite nuclear matter was proposed \cite{Margueron18a}. This technique 
possesses the flexibility of agnostic approaches within the functional forms allowed by
the hypothesis of beta-equilibrium in a matter composed of neutrons and protons. 
As such, it can also take care of the experimental constraints coming from laboratory 
by employing semi-classical approximations such as 
the extended Thomas-Fermi approach \cite{Aymard16a, Aymard16b, Chatterjee17}. 
This approach also provides further advantage of treating the 
crust and core of neutron star matter  in unison 
\cite{Dinh-Thi21, DinhThi21a, DinhThi21b}, importance of which has been 
quantified in recent times \cite{Suleiman21, Compose22}.
 
The present study is intended towards exploring in details the impact of 
recent measurement of neutron skin in $^{208}$Pb and $^{48}$Ca in 
Jefferson lab on the knowledge of the nuclear EoS, in 
light of the existing constraints from astrophysical observations.
To this aim, we have employed the nuclear meta-model for the EoS, and put to use further 
an analytical version of the extended Thomas-Fermi (ETF) model to calculate 
few ground state properties of finite nuclei \textit{e.g.} binding 
energy, charge radii, and neutron skin \cite{Aymard16a, Aymard16b, 
Chatterjee17}. We performed a full Bayesian study for static astrophysical 
observables as well as ground state finite nuclear properties. 

The paper is organized as follows. In Section \ref{formalism} we outline 
briefly the meta-modeling of EoS and excerpts of the analytical ETF model 
developed by Aymard \textit{et. al.} in Ref. \cite{Aymard16a, Aymard16b}. 
In section \ref{bayesian-cal} we provide the details of the Bayesian analysis 
employed in this calculation. Our results are discussed in Section  
\ref{results}. The concluding remarks are drawn in Section \ref{conclusions}.

\section{Formalism}\label{formalism}
\subsection{EoS metamodel in $\beta$-equilibrium}
The energy per particle of infinite homogeneous nuclear matter composed of neutrons and protons 
at density $n_n$ and $n_p$, respectively, is written as\cite{Margueron18a},
\begin{eqnarray}\label{meta-def}
e(n,\delta)=C_{kin}\sum_{q=n,p}\frac{n_q^{5/3}}{m^\star_q(n,\delta)} +U_0(n)+U_{sym}(n)\delta^2, 
\end{eqnarray}
where $n=n_n+n_p$ is the total density, $\delta=(n_n-n_p)/n$ is the isospin 
asymmetry, and $C_{kin}=3(3\pi^2\hbar^3)^{2/3}/10$. The first term accounts 
for the zero point nuclear motion, and the dominant density dependence arising from the non-locality of the 
effective interaction, while the density dependence associated to the 
symmetric $U_0(n)$ and asymmetric $U_{sym}(n)$ part of the local nuclear 
potential is given by an agnostic Taylor expansion around the 
saturation point of symmetric matter $n_{sat}$ as
\begin{eqnarray}
\label{metapot}
U_{0,sym}(n)=\sum_{k=0}^4 \frac{(v_k)_{0,sym}}{k!}x^k u^{N = 4}_{k}(x),
\end{eqnarray}
where $x = (n - n_{sat})/(3n_{sat})$, and $u^{N}_{k}(x) = 1 - (-3x)^{N+1-k}
\exp(-b(1+3x))$ with $b$  a correction  ensuring the convergence 
at the zero-density limit. The density dependence of the effective 
masses $m^\star_q$ in Eq. (\ref{meta-def}) is governed by two parameters 
$\kappa_{sat}$ and $\kappa_{sym}$~\cite{Margueron18a} that are physically 
connected to the empirical value of the isoscalar effective mass 
$m_{sat}^{\star}$ and its isovector splitting $\Delta m^{\star}/m$, both 
known experimentally, albeit with fair amount of uncertainties 
\cite{Dutra12, Dutra14, Li15b, Zhang16, Kong17, Mondal18, Mondal17a}.

The coefficients $(v_k)_{0,sym}$ can be expressed solely in terms of the 
so-called nuclear matter empirical parameters (NMPs). These 
correspond to different coefficients of Taylor's expansion in density 
around the saturation point $n_{sat}$ of the symmetric matter (SNM) energy 
$e_0(n)\equiv e(n,\delta)|_{\delta=0}$ and symmetry energy $e_{sym}(n)
\equiv {1\over 2}\frac{\partial^2 e}{\partial \delta^2}|_{\delta=0}$.
Retaining up to fourth order, 
in $e_0(n)$ these are energy per particle $E_{sat}$,
incompressibility $K_{sat}$, skewness $Q_{sat}$ and stiffness $Z_{sat}$; in 
$e_{sym}(n)$ they are symmetry energy $E_{sym}$, symmetry slope
$L_{sym}$, symmetry incompressibility $K_{sym}$, symmetry skewness $Q_{sym}$ and
symmetry stiffness $Z_{sym}$.  
In both isoscalar and isovector sector, the NMPs are relatively tightly 
constrained by experiments up to order 2, thus allowing educated priors for 
the Bayesian treatment.

For the computation of the cold neutron star EoS, the composition of matter 
is determined by solving the coupled equations of nucleonic $\beta$-equilibrium,
\begin{eqnarray}
	\label{betaeq}
	&\mu_n(n,\delta_\beta)-\mu_p(n,\delta_\beta)=\mu_e(n,\delta_\beta),&\\
	&2\left.\frac{\partial e(n,\delta)}{\partial \delta}\right|_n=
	\mu_e(n,\delta)-(m_{n}-m_{p}),&\\
	&\mu_e = C_e \left[\gamma_r
	(1+6x_r^2)+\frac{x_r^2(2x_r^2+1)}{\gamma_r}-\frac{1}{\gamma_r}\right],&
\end{eqnarray}
where $C_e=\frac{(m_e)^3}{8(3\pi^2n_e)^{2/3}(\hbar c)^2}$, $x_r = 
\frac{\hbar c (3\pi^2 n_e)^{1/3}}{m_e}$, and $\gamma_r = \sqrt{1+x_r^2}$; 
$\mu_{n,p,e}$ and $m_{n,p,e}$ are the chemical potentials and free masses of neutron, 
proton and electron, respectively; and $n_e$ is the density of electrons. Muons appear in the 
system spontaneously when the lepton chemical potential $\mu_e$  exceeds the muon free mass
$m_{\mu}$, and their density is fixed by $n_{\mu}=n_p-n_e$ in the global equilibrium 
condition $\mu_{\mu}={\mu}_e$.
Once we get the composition solving the beta equilibrium equations, the baryonic pressure 
 can be calculated as, 
\begin{eqnarray}
	\label{metapres}
p_{bar}(n,\delta)=n^2 \frac{\partial e(n,\delta)}{\partial n}.
\end{eqnarray}

In the neutron star crust, the meta-modeling is extended to treat finite nuclei 
in the compressible liquid drop model (CLDM) approximation \cite{Carreau19}. 
To describe a spherical nucleus of mass number $A$, charge $Z$, bulk density 
$n_i$ and radius $r_N$ in a spherical Wigner-Seitz (WS) cell of radius $r_{WS}$, the bulk 
energy $E_{bulk}=A e(n_i,1-2Z/A)$ is complemented with  Coulomb, surface, 
and curvature terms. The Coulomb energy is given by:
\begin{equation}
{E}_{Coul}  = \frac{8}{3}\left (\pi eZ n_i\right )^2r_N^5\eta_{\rm Coul}
	\left(\frac{r_N}{r_{WS}}\right), \label{eq:Ecoul}
\end{equation}
where $e$ is the elementary charge, and the function $\eta_{\rm Coul}(x)$ 
accounting for the electron screening is written as  
\begin{equation}
    \eta_{\rm Coul}(x) =\frac{1}{5}\left [ x^3+ 2 \left ( 1- \frac{3}{2}x \right ) 
	\right ]. \label{eq:eta_Coul}
\end{equation}
The surface and curvature energies are expressed as:  
\begin{equation}
E_{\rm {surf}} + E_{\rm {curv}} =4\pi r_N^2\left ( \sigma_s(Z/A) +
	\frac{2\sigma_c(Z/A)}{r_N}\right ) , \label{eq:interface}   
\end{equation}
where $\sigma_s$  and $\sigma_c$ are the surface and curvature tensions, 
with an isospin dependence based on the behavior of Thomas-Fermi 
calculations at extreme isospin asymmetries \cite{Ravenhall83}
\begin{eqnarray}
\sigma_s(x)&=&\sigma_0\frac{2^{4}+b_s}{x^{-3}+b_s+(1-x)^{-3}} \ , \label{eq:surface} \\
\sigma_c(x)&=&5.5 \, \sigma_s(x) \frac{\sigma_{0,c}}{\sigma_0}(\beta-x)\ . \label{eq:curvature} \ 
\end{eqnarray}

For a set of bulk parameters appearing in the energy functional of Eq.(\ref{meta-def}), 
the bulk energy of any nucleus $(A,Z)$ in the vacuum is given by 
$E_{bulk}^{vac}=A e(n_i^{vac},1-2Z/A)$, where $n_i^{vac}$ is the solution of the equation 
$\partial e(n, 1-2Z/A)/\partial n=0$.
The parameters corresponding to the surface and curvature terms $\sigma_0$, 
$\sigma_{0,c}$, $b_s$ and $\beta$ are then optimized on the AME2016 mass 
table \cite{Wang17,DinhThi21a,DinhThi21b}. As a consequence, the physical 
correlation between bulk and surface parameters embedded in the empirical 
value of the nuclear masses is insured. The crustal EoS is finally 
determined by minimizing the energy of the WS cell with 
respect to the parameters defining the crustal composition ($A,Z,n_i,
r_N,r_{WS}$, and the dripped neutron density $n_g$), as in refs. 
\cite{Carreau19,DinhThi21a,DinhThi21b}.

The CLDM description of the ground state of finite nuclei misses shell 
effects and specific properties of the effective nucleon-nucleon 
interaction such as spin-orbit coupling and tensor terms. Because of 
that, its predictive power is obviously quite limited. However, if the 
parameters are fitted on a large sample of nuclear masses it was 
recently shown that the CLDM energy compares reasonably well with more 
microscopic extended Thomas Fermi (ETF) approaches \cite{Furtado20,Grams22}. 
Moreover, though the composition of the crust is not the same as the 
one obtained in full Hartree-Fock-Bogoliubov (HFB) throry, 
the crustal EoS is very well reproduced 
\cite{Carreau19a}. For this reason, we consider that the CLDM approach 
is sophisticated enough to realistically predict the NS crustal EoS. 
The improved treatment that we adopt to predict the ground state observables, notably the skin, is described in the next section.

\subsection{Analytic Extended Thomas-Fermi method for nuclei}
As far as ground state nuclear observables such as radii and skins are concerned,
the CLDM approximation is not adequate and it is important to account for the full neutron 
and proton density profiles $n_n(r)$ and $n_p(r)$. Full HFB calculations including nuclear deformation and time-odd terms are in principle necessary for the purpose, and efficient numerical codes start to be available \cite{Scamps21}. 
However, these approaches being numerically 
too expensive for a large Bayesian analysis, we resort to the ETF approximation, 
that was successfully compared with experimental data on binding energies and radii 
since many decades, see refs.\cite{Chatterjee17,Bhagwat21} for recent works, and 
references therein. Another advantage of the ETF method is that the integral expressions 
giving the nuclear energy and radii can be analytically calculated \cite{Aymard16a,
Aymard16b} within some approximations that are well justified for nuclei not too far 
from stability as the ones considered in this section. 
This produces an analytical ETF mass formula that is ideally suited for the Bayesian 
analysis of the correlations between the observables and the EoS. The main aspects of 
the model are briefly recalled in this section, for more details see 
refs.\cite{Aymard16a,Aymard16b}.

We start from the expression of the strong interaction part of the nuclear binding 
for a spherical nucleus in the ETF approximation: 
\begin{eqnarray}\label{enuc}
	E_{nuc}=4\pi \int_0^{\infty}dr r^2 \mathcal{H}_{ETF}[n_n(r),n_p(r)].
\end{eqnarray}
The ETF functional at the second order in $\hbar$ is given by, 
\begin{eqnarray}\label{hetf}
	\mathcal{H}_{ETF}[n_n(r),n_p(r)]&=&e(n_n,n_p)n_0+\sum_{q=n,p}\frac{\hbar^2}{2m_q^{\star}}
	\tau_{2q}\nonumber\\&& +C_{fin}(\boldsymbol{\nabla} n_0)^2.
	\label{ETF}
\end{eqnarray}
Here, $e(n_n,n_p)$ comes directly from the meta-model energy functional 
of Eq. (\ref{meta-def}), evaluated at the local densities. 
In Eq. (\ref{ETF}), the local and non-local $\hbar^2$ corrections 
$\tau_{2q}=\tau_{2q}^l+\tau_{2q}^{nl}$ are given by, 
\begin{eqnarray}
	\tau_{2q}^l&=&\frac{1}{36}\frac{(\boldsymbol\nabla n_q)^2}{n_q}+\frac{1}{3}
	\Delta n_q,\nonumber\\
	\tau_{2q}^{nl}&=&\frac{1}{6}\frac{\boldsymbol\nabla n_q \boldsymbol\nabla f_q}{f_q}
	+\frac{1}{6}n_0\frac{\Delta f_q}{f_q}-\frac{1}{12} n_q\left(\frac{\boldsymbol
	\nabla f_q}{f_q}\right)^2,  
\end{eqnarray}
where $f_q=\frac{m}{m^{\star}_q}$, with $m$ the bare nucleon mass and $m^{\star}_q$, $q=n,p$, 
giving the effective masses, already present in the zero order $\hbar$ expression Eq. (\ref{meta-def})
. $C_{fin}$ is an extra parameter controlling the dominant gradient correction to the local functional.
One may observe that realistic microscopic functionals contain more couplings 
related to gradient terms, notably at least the spin-orbit term. However, the associated parameters are 
strongly correlated. In order to pin down the EoS dependence, it was suggested that it 
might be sufficient to introduce a single effective $C_{fin}$ parameter in the isoscalar 
sector \cite{Chatterjee17}. We expect that the presence of 
extra couplings in finite nuclei with respect to the simplified case of homogeneous 
nuclear matter, will weaken the correlations between properties of finite nuclei and the 
nuclear EoS. Our choice of allowing for a unique gradient parameter will therefore 
give upper limits for those correlations. Anticipating our results, we will show that those correlations are weak, meaning 
that such upper limits are going to be quite significant.
Concerning the isovector sector, an extra 
parameter $Q$ is introduced below directly in the parametrization of the density 
profiles, to effectively account for isospin dependent gradient terms.

The densities in the ETF integral are commonly employed as Fermi 
functions to perform the integration analytically as,
\begin{eqnarray}
	n_q(r)=n_{bulk,q}F_q(r), \nonumber\\ F_q(r)=\frac{1}{1+e^{(r-R_q)/a_q}},\nonumber\\ 
	n_{bulk,q}=n_{bulk}(\delta)\frac{1\pm\delta}{2}.
\end{eqnarray}
Here, $R_q$ and $a_q$ are the radius and diffuseness parameters of
the nucleon density profiles. The bulk density $n_{bulk}(\delta)$ and bulk asymmetry $\delta$ associated to a nucleus 
with proton number $Z$ and neutron number $N$ are determined by solving the 
following equations self-consistently \cite{Aymard16b}, 
\begin{eqnarray}
	\delta&=&\frac{\frac{N-Z}{A}+\frac{3a_cZ^2}{8QA^{5/3}}}{1+\frac{9E_{sym}}{4QA^{1/3}}},\\
	n_{bulk}(\delta)&=&n_{sat}\left(1-\frac{3L_{sym}\delta^2}{K_{sat}+K_{sym}
	\delta^2}\right),\\
	a_c &=& \frac{3e^2}{20\pi\varepsilon_0 r_{bulk}(\delta)}.
\end{eqnarray}
In the equations above,  $Q$ is the so-called surface stiffness, linked to the average 
distance between proton and neutron surfaces \cite{Warda09}, and $a_c$ is the Coulomb 
parameter with $r_{bulk}=(3/4\pi n_{bulk})^{1/3}$. With the approximation $a_n=a_p=a$, 
the diffuseness of the density distribution can be variationally obtained as \cite{Aymard16b}, 
\begin{eqnarray}
	&&a^2(A,\delta) = \frac{\mathcal{C}_{surf}^{NL}(\delta)}{\mathcal{C}_{surf}^{L}(\delta)}+
	\Delta R_{HS}(A,\delta)\sqrt{\frac{\pi}{\left(1-\frac{K_{1/2}}{18J_{1/2}}\right)}}\nonumber\\
	&&\ \ \ \ \times\frac{n_{sat}}{n_{bulk}(\delta)}\frac{3J_{1/2}}{\mathcal{C}_{surf}^{L}(\delta)}
	\sqrt{\frac{\mathcal{C}_{surf}^{NL0}}{\mathcal{C}_{surf}^{L0}}}
	\left(\delta-\delta^2\right). \label{eq:diffuseness}
\end{eqnarray}
In this equation, $J_{1/2}$ and $K_{1/2}$ are the symmetry energy coefficients of order 
0 and 2, respectively, calculated at the density $n=n_{sat}/2$; $J_{1/2}=e_{sym}(n_{sat}/2)$, 
$K_{1/2}=9(n_{sat}/2)^2\partial^2 e_{sym}/\partial n^2|_{n_{sat}/2}$. The coefficients 
${\cal{C}}_{surf}^{L,NL}(\delta)$ and ${\cal{C}}_{surf}^{L0,NL0}$ $\equiv$ 
${\cal{C}}_{surf,curv,ind}^{L,NL}(\delta=0)$, depend both on the local interaction 
parameters $(v_k)_{0,sym}$ of Eq.(\ref{meta-def}) and on the effective masses, and 
their explicit expressions are given in  Refs. \cite{Aymard16a,Aymard16b}.
Finally, the hard sphere radii of the total ($R_{HS}$) and proton ($R_{HS,p}$) distribution  are introduced as
\begin{eqnarray}
	\Delta R_{HS} &=& R_{HS}-R_{HS,p} \nonumber\\
	&=& r_{bulk}(\delta)A^{1/3} - 
	r_{bulk,p}(\delta)Z^{1/3}\nonumber\\
	&=&\left(\frac{3}{4\pi}\right)^{\frac{1}{3}}\left[
	\left(\frac{A}{n_{bulk}(\delta)}\right)^{\frac{1}{3}}-
		\left(\frac{Z}{n_{bulk,p}(\delta)}\right)^{\frac{1}{3}}\right].\nonumber\\ 
\end{eqnarray}
The analytical expression for the diffuseness of the density profile in Eq. 
(\ref{eq:diffuseness}) allows one to compute the nuclear mass by direct integration of Eq. 
(\ref{enuc}), with the addition  of a Coulomb term, 
\begin{equation}
    M(A,Z)= Nm_n+ Zm_p + E_{nuc}(A,Z)+a_c \frac{Z^2}{A^{1/3}} .
\end{equation}
Moreover, once the mean square radii are given as
\begin{eqnarray}
	\left<r_{q}^2\right>=\frac{3}{5}R_{HS,q}^2\left(1+\frac{5\pi^2a^2}{6R_{HS,q}^2}\right)^2,
\end{eqnarray}
the neutron skin $\Delta r_{np}$ and the charge radii $R_{ch}$ are obtained from, 
\begin{eqnarray}
	\Delta r_{np}=\sqrt{\left<r_{n}^2\right>}-\sqrt{\left<r_{p}^2\right>},\\ 
	R_{ch}=\left[{\left<r_{p}^2\right>}+S_p^2\right]^{\frac{1}{2}}.
\end{eqnarray}
The correction term $S_p$ corresponds to the internal charge distribution of protons, 
which is taken as 0.8 fm \cite{Buchinger94,Patyk99}. This quasi-analytic ETF method to calculate the gross properties of 
nuclei will be referred as ``{\it aETF}" henceforth.

\section{Bayesian Analysis}\label{bayesian-cal}
We perform a Bayesian analysis for the different properties of nuclei using 
the \textit{aETF} method described in the previous section following Ref. 
\cite{Aymard16a, Aymard16b}, as well as properties of neutron star following 
the metamodelling technique of Ref. \cite{Dinh-Thi21}. 
First of all, the nuclear parameters that can be largely varied to build 
our prior EoS model, are the set of 12 NMPs corresponding to infinite nuclear 
matter, namely: $n_{sat}$, $E_{sat,sym}$, $L_{sym}$, $K_{sat,sym}$, $Q_{sat,sym}$, 
$Z_{sat,sym}$, $\kappa_{sat,sym}$. It was shown in Ref. \cite{Margueron18a} that, 
if the Taylor expansion is truncated at the order $N=4$, to reproduce precisely 
any arbitrary nuclear model in a large density domain, it is necessary 
to consider different values for the third and fourth order NMPs \textit{i.e.} 
$Q_{sat,sym}$ and $Z_{sat,sym}$ below and above $n_{sat}$, respectively. 
For this reason, we sample those NMPs as separate parameters 
below and above saturation density, leading to a total number of 16 independent NMPs. 
In this way, we make sure that the low density behavior of the EoS, as imposed 
by the observables sensitive to subsaturation density, does not impose any spurious 
correlation to the high density regime, where higher order derivatives start to play a role, 
and additionally different degrees of freedom may pop 
out \cite{Biswas21}. We remark that this procedure does not include any discontinuity in the 
pressure nor in the sound speed. We denote henceforth the high order parameters 
above saturation with an asterisk mark ($Q_{sat,sym}^{\star}$ and 
$Z_{sat,sym}^{\star}$). The ranges for these NMPs used in the present 
work are provided in tabulated form in the supplemental material. 

\begin{figure}[]{}
{\includegraphics[height=2.8in,width=2.5in,angle=-90]{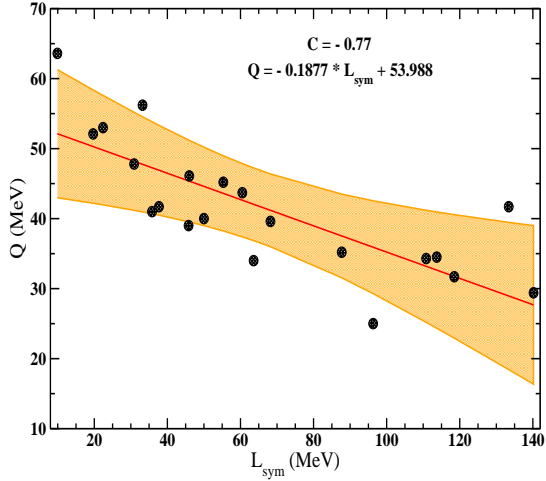}}
\caption{\label{qlcorr}
(Color online) Surface stiffness $Q$ plotted as a function of symmetry 
	slope $L_{sym}$ for the models given in Ref. \cite{Warda09} along 
	with few more in Ref. \cite{Farine81}. The fitted straight line 
	is also depicted in red accompanied by the Pearson correlation 
	coefficient $C$.}
\end{figure}

Apart from sampling the NMPs ruling the behavior of homogeneous  
nuclear matter with independent flat distributions, one needs further parameters $C_{fin}$ and 
$Q$ to calculate the properties of nuclei entering in Eqs. (\ref{enuc}) and (\ref{hetf}). 
We sample $C_{fin}$ between 40 and 80 MeV in a flat distribution following the 
optimized value obtained in Ref. \cite{Chatterjee17}. Concerning the 
surface stiffness $Q$, in principle this parameter can be extracted from semi-infinite 
nuclear matter slab calculations in the Thomas Fermi or ETF approximation \cite{Warda09}. 
This means that we expect it to be somehow correlated to the bulk parameters, 
to the gradient term $C_{fin}$ and possibly to other gradient terms neglected 
in the present study. The importance of this correlation clearly depends on the 
detailed expressions assumed for the functional. For this reason, we have followed 
two different procedures corresponding to two extreme hypotheses on the degree of 
independence of $Q$ from the rest of the parameter set. In the first, we sample 
$Q$ fully agnostically within a  flat probability distribution 
between 5 and 70 following the range of its values obtained in the 
literature \cite{Farine81, Warda09}. In a complementary method, we make 
use of the fact that $Q$ was found to be linearly correlated 
with symmetry slope parameter $L_{sym}$ across many popularly used mean 
field models in the literature. We sample $Q$ according to its correlation 
with $L_{sym}$ as depicted in Fig. \ref{qlcorr}. The Pearson correlation 
coefficient is $-0.77$, which clearly points towards some ambiguity. To 
account for this deviation from the absolute correlation, we obtained the 
99.9\% confidence band of $Q$, from where we sampled $Q$ randomly within its 
extremities for a given value of $L_{sym}$.

All in all, we performed our analysis with $N_p=19$ parameters in $Q$-$L_{sym}$ 
uncorrelated version, and with $N_p=18$ parameters where further aid 
is taken from Fig. \ref{qlcorr}. The parameter set is collectively named 
as $\mathbf X\equiv \{X_k,k=1,\dots N_p\}$.
The prior distribution $P_{prior}(\mathbf X)=\prod_{k=1}^{N_p}P_k({X_k})$ of 
the \textit{a-priori} uncorrelated parameter set is obtained 
with flat distributions $P_k({X_k})$, with intervals $\{ X_k^{min},X_k^{max}\}$  that 
are detailed in the supplemental material. The high density third and fourth order NMPS $Q_{sat,sym}^{\star}$ and 
$Z_{sat,sym}^{\star}$ were populated with the same limit as their 
non-asterisked partners. 
Posterior distributions are subsequently obtained using different 
physical filters, as outlined in the next section.


\subsection*{Filters}\label{formalism_filter}
1. {\it AME+R$_{ch}$:} The standard likelihood expression for this filter is given by,
\begin{equation}
P_{\text{AME+R}_{ch}}(\mathbf X) \propto \omega_0 e^{-\chi_{AME}^2({\bf X})/2}e^{-\chi_{Rch}^2({\bf X})/2} 
\prod_{i=1}^{N_p} P_i({X_i}),
\label{P_prior}
\end{equation}
where $\omega_0 = 0$ or 1 depending on the meaningful production of nuclear 
masses and charge radii having a meaningful solution for the bulk asymmetry $\delta$ 
and the diffuseness of the density profile $a$; and $P({\mathbf X})$ 
corresponds to the flat prior distribution of 
different meta model and \textit{aETF} parameters. The objective 
function for the AME2016 mass table and the experimental 
charge radii for a few spherical nuclei appearing above are defined by 
\begin{eqnarray}
	\chi^2_{AME}({\bf X})&=&\frac{1}{N_1}\sum_n \frac{\left 
	( M_{ETF}^{(n)}({\bf X})-M_{AME}^{(n)}\right )^2}
	{\sigma_{BE}^2},\text{ and} \nonumber\\
	\chi^2_{Rch}({\bf X})&=&\frac{1}{N_2}\sum_n \frac{\left
	( R_{ch(n)}^{ETF}({\bf X})-R_{ch(n)}^{exp}\right )^2}
        {\sigma_{ch}^2}.
        \label{eq_chi2}
\end{eqnarray}
Here, we have $N_1=2408$ and $N_2=9$. The $\sigma_{BE}$ are chosen 
as 1\% of the corresponding mass and $\sigma_{ch}^2=0.02$ fm for 7 
nuclei and 0.1 for 2 nuclei. The  specifics of the charge 
radii \cite{Angeli13} used in the present work are detailed in the supplemental material. 

Since a satisfactory reproduction of mass and radii is a necessary condition for using
a nuclear functional in the reproduction of more sophisticated observables such as the nuclear skin,
the successive filters are all applied on top of AME+R$_{ch}$, such that this first posterior plays the role of a
prior informed by nuclear experiments on ground state properties.
 The posterior probability distributions of the set 
$\bf X$ of  EoS and \textit{aETF} parameters for other filters are conditioned by 
likelihood models of the different observations and constraints $\mathbf c $ with 
normalizing constant $\mathcal N$ as:
\begin{equation}
        P({\mathbf X}|{\mathbf c})=\mathcal N   P_{\text{AME+R}_{ch}}({\mathbf X}) \prod_k P(c_k|{\mathbf X}). 
\end{equation}
The corresponding probability distributions for the observables $Y({\bf X})$ 
are obtained by an overall marginalization through the range of values of 
parameters $\bf X$ between ${\bf X}_{min}$ and ${\bf X}_{max}$ according to
\begin{equation}
        P(Y|{\mathbf c})=\prod_{k=1}^N\int_{X_k^{min}}^{X_k^{max}}dX_k \, 
	P({\mathbf X}|{\mathbf c}) \delta\left (Y-Y({\mathbf X})\right ).
\end{equation}

2. {\it	$\chi$-EFT:} The next filter we apply on our nuclear physics informed prior is the 
constraints on symmetric nuclear matter (SNM) and pure neutron matter 
(PNM) at low densities from 0.02 - 0.2 fm$^{-3}$ obtained by 
theoretical calculations from the chiral effective field theory 
({\it $\chi$-EFT}) \cite{Drischler16}. The probability of the posterior 
distribution can be outlined as
\begin{equation}
P_{\chi\text{-}EFT}(\mathbf X) \propto \omega_{\chi\text{-}EFT}(\mathbf X) 
	P_{\text{AME+R}_{ch}}(\mathbf X),
\label{P_LD}
\end{equation}
where, $\omega_{\chi\text{-}EFT}=0$ or 1, depending on whether they pass 
through the area predicted by the {\it $\chi$-EFT} calculations. 
This theoretical band is interpreted as a $1-\sigma$ uncertainty,  
and for this reason a 5\% extension is added on the edges. 

3. {\it Astro}: The high density part of the nuclear matter is 
known to be quite sensitive to the constraint on the observed maximum mass 
of neutron star obtained by measuring Shapiro delay \cite{Antoniadis13, 
Demorest10}. We outlined the likelihood probability of a model $\bf X$ 
by taking a cumulative Gaussian distribution function for the observed 
maximum mass $M_{max}$ as depicted in Ref. \cite{Antoniadis13} with mean 
at $2.01 M_{\odot}$ with variance $0.04 M_{\odot}$ as, 
\begin{equation}
P(M_{max}|{\mathbf X}) =\frac{1}{0.04\sqrt{2\pi}} \int_{0}^{M_{max}({\bf X})
/M_{\odot}}e^{-\frac{(x-2.01)^2}{2\times 0.04^2}}dx.
\label{mmax}
\end{equation}

Some effects are also imparted by the data on joint tidal deformability 
$\tilde\Lambda$  of the GW170817 event, which is defined as
\begin{equation}
\tilde{\Lambda}=\frac{16}{13}\frac{(m_1+12m_2)m_1^4\Lambda_1+(m_2+12m_1)m_2^4\Lambda_2}{(m_1+m_2)^5},
\label{tidal}
\end{equation}
where, $m_1, m_2$ are the masses of the merging NS system and $\Lambda_1, 
\Lambda_2$ are their respective tidal deformabilities, which are connected 
to the mass $M$ and radius $R$ of the corresponding system as $\Lambda
\propto (R/M)^5$. The GW170817 event gives direct constraint on the 
$\tilde\Lambda$ and mass-ratio $q$ for an event with chirp mass $\mathcal{M}_{chirp} 
=1.186\pm0.001 M_{\odot}$ in a three dimensional posterior probability 
distribution. As $\mathcal{M}_{chirp}$ is  very precisely measured, we defined 
the likelihood as
\begin{equation}
P(LVC|{\mathbf X})= \sum_{i} P_{LVC}(\tilde{\Lambda}(q^{(i)}), q^{(i)}),
\label{LVC}
\end{equation}
where, $P_{LVC}(\tilde{\Lambda}(q), q)$ is the approximated two dimensional 
posterior probability from GW170817 event obtained by LIGO-Virgo collaboration 
(LVC) \cite{Abbott19}, which we interpolated for each $\bf X$ in our calculation 
by sampling $q \in \left[0.73, 1.00\right]$ and calculating the corresponding 
masses, making the approximation $\mathcal{M}_{chirp}=1.186$ for all samples, as
\begin{equation}
\mathcal{M}_{chirp} = \frac{(m_1m_2)^{3/5}}{(m_1+m_2)^{1/5}} =\frac{q^{3/5}m_1}{(1+q)^{1/5}}, 
\label{chirpmass}
\end{equation}
and the corresponding $\tilde\Lambda$ from Eq. (\ref{tidal}). 

Finally, the posterior probability of this distribution is written as:
\begin{equation}
P_{astro}({\bf X}) \propto \omega_{astro}P(M_{max}|{\mathbf X})
	P(LVC|{\mathbf X})P_{\text{AME+R}_{ch}}(\mathbf X).
\label{P_HDLVC}
\end{equation}
Here, $\omega_{astro}$ is a pass-band type filter  similar to 
$\omega_{\chi\text{-}EFT}$ in Eq. (\ref{P_LD}). This filter  eliminates models which violate causality
 or thermodynamic stability, or exhibit negative symmetry energy at densities lower than 
the central density of the highest NS mass $M_{max}$ for the corresponding sample, or 
leading to a non-convergent solution for the variational equations of the crust.
We do not explicitly include the information from NICER observations 
\cite{Riley19,Riley21,Miller19,Miller21}, because it was shown in ref.\cite{Dinh-Thi21} 
that the present uncertainties are such that these observations
do not add major constraints on the metamodelling parameters, if the model is already 
informed by the GW170817 gravitational wave data.

\begin{figure}[]{}
{\includegraphics[height=2.8in,width=2.8in]{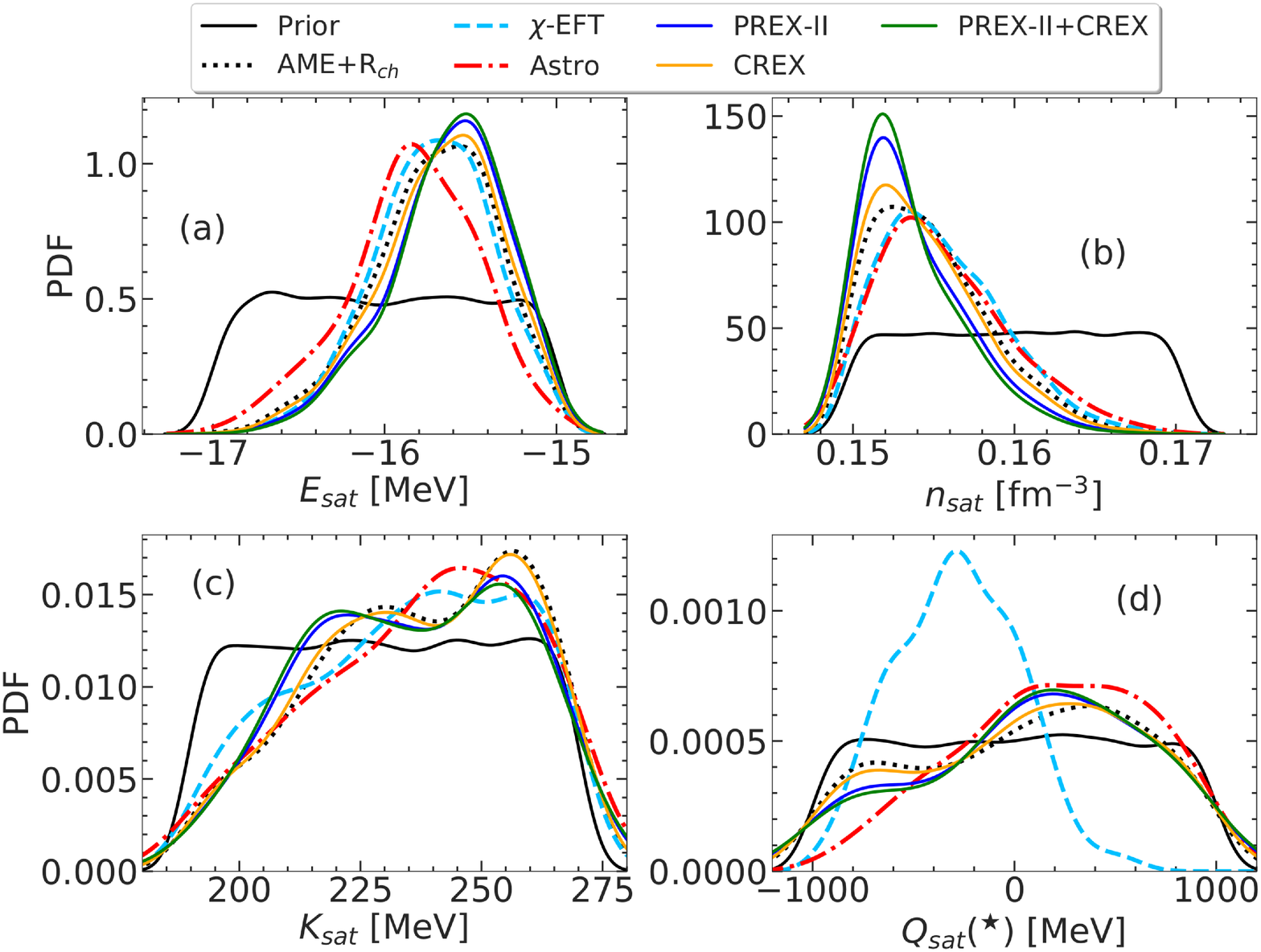}}
\caption{\label{satnm}
(Color online) Posterior probability distributions of different empirical 
	parameters corresponding to the symmetric nuclear matter obtained with 
	different filters described in Section \ref{formalism_filter}. 
	}
\end{figure}
4. {\it PREX-II+CREX}: The posterior probability of this distribution is written as:
\begin{equation}
P_{PREX+CREX}(\mathbf X) \propto e^{-\chi_{skin}^2({\bf X})/2} P_{\text{AME+R}_{ch}}({\mathbf X}),
\label{P_skin}
\end{equation}
where the cost function for the neutron skin measurement is defined by 
\begin{eqnarray}
	\chi^2_{PREX+CREX}({\bf X})&=&\frac{\left (\Delta r_{np}^{208}({\bf X})-\Delta r_{np}^{\text{PREX-II}}\right )^2}
	{(0.07)^2}, \nonumber\\
	&+&\frac{\left (\Delta r_{np}^{48}({\bf X})-\Delta r_{np}^{\text{CREX}}\right )^2}
        {(0.05)^2}.
        \label{eq_chi2skin}
\end{eqnarray}
Here, $\Delta r_{np}^{208}({\bf X})$ and $\Delta r_{np}^{48}({\bf X})$ are the 
values of $\Delta r_{np}$ in fm corresponding to $^{208}$Pb and $^{48}$Ca nuclei, 
calculated with the parameter set $\bf X$. The values of $\Delta r_{np}^{\text{PREX-II}}=0.283$ fm 
$\Delta r_{np}^{\text{CREX}}=0.121$ are taken from Refs. \cite{Adhikari21, Adhikari22}, 
respectively. It is important to point out here that, we have applied these 
filters from PREX-II and CREX separately as well, to observe their individual 
impact on different observables of interest.

\section{Results}\label{results}
\begin{figure}[]{}
{\includegraphics[height=2.8in,width=2.8in]{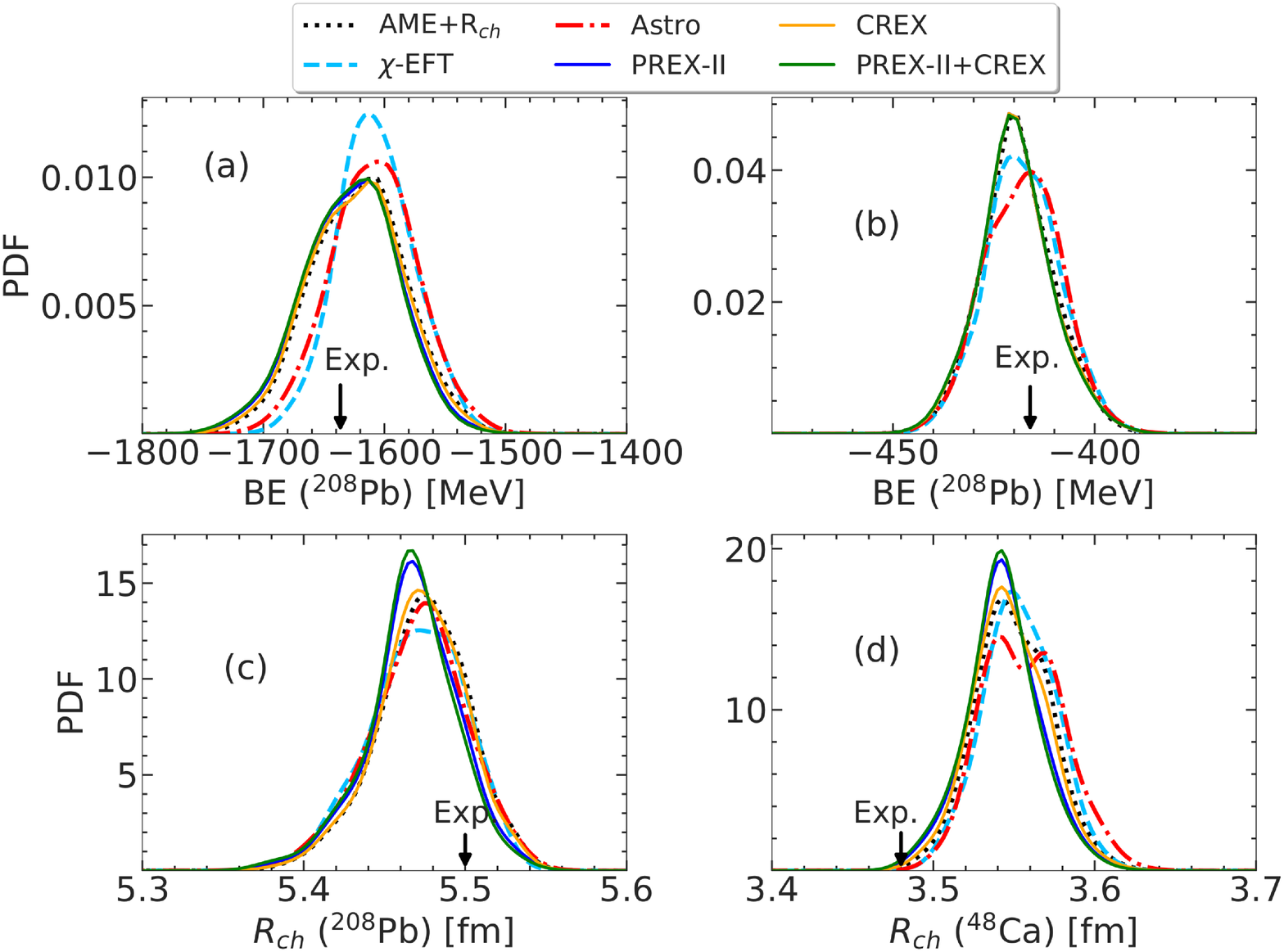}}
\caption{\label{bech48208}
(Color online) Posterior probability distributions of binding energies (upper panels) and charge 
	radii (lower panels) of $^{208}$Pb (left) and $^{48}$Ca (right) 
	nuclei obtained with the \textit{aETF} method corresponding 
	to the same filters as in Fig. \ref{satnm}.
        }
\end{figure}

In Fig. \ref{satnm} we plot the one dimensional probability distributions 
of different empirical parameters corresponding to symmetric matter 
using the filters described in Sec. \ref{formalism_filter}, 
and the sampling of $Q$ parameter following the correlation systematics of 
Fig. \ref{qlcorr}. The behavior is almost indistinguishable from the one 
obtained with the other method of $Q$ sampling.  The latter is thus only presented 
in the supplemental material. The separate effects of PREX-II and CREX 
are also displayed in the figure. Since \textit{AME+R$_{ch}$} is informed by nuclear data, 
one can already observe peaks in the distributions for $E_{sat}$ and 
$n_{sat}$ around $-16.1$ MeV and $0.154$ fm$^{-3}$, respectively, and the 
distributions of those parameters get only marginally affected by the subsequent 
information from nuclear theory, skin data, or astrophysical observations. 
The effect of the mass constraint gets diluted in $K_{sat}$ and degrades 
further in $Q_{sat}$. We do not include results for the fourth-order 
parameters $Z_{sat}$, $Z_{sat}^{\star}$ because they have very large 
uncertainties and very little impact from the different constraints. 
The same applies to $Z_{sym}$, $Z_{sym}^{\star}$ in the symmetry 
energy sector, to be shown in the upcoming figures. The
peak in $n_{sat}$ at the level of mass-informed prior was not 
observed in  a study performed with a similar technique in
Ref. \cite{Dinh-Thi21}. Anticipating the results from the correlation study 
below, this occurs here because of the inclusion of the constraints 
on charge radii using the \textit{aETF} method in the present calculation, while only 
binding energy constraints were used in Ref. \cite{Dinh-Thi21}. One can also observe that 
the data on neutron skins from CREX and PREX-II have no impact on these isoscalar 
properties, which is expected. 

In Fig. \ref{satnm}(d), the quantity represented is the skewness $Q_{sat}$ 
below saturation for all distributions except \textit{Astro}. For 
that, we have plotted the effective supersaturation skewness parameter 
$Q_{sat}^{\star}$, which is constrained better by this filter, 
particularly being sensitive to the high density behavior of the EoS. 
The distinct peak appearing for $Q_{sat}$ in \textit{$\chi$-EFT}, 
which is located at a different place than that in \textit{Astro}, 
does not therefore point towards a possible tension between nuclear theory and astrophysical 
measurements. It rather emphasizes the fact that the high density EoS behavior 
cannot be extrapolated from the sub-saturation EoS, even in the conservative hypothesis that 
no exotic degrees of freedom appear at high density. Indeed, higher order terms in 
the Taylor expansion become dominant at high density, and are here effectively 
summed up as a $Q_{sat}^{\star}$ contribution. We will further comment this point in the correlation study below.
\begin{figure}[]{}
{\includegraphics[height=2.8in,width=2.8in]{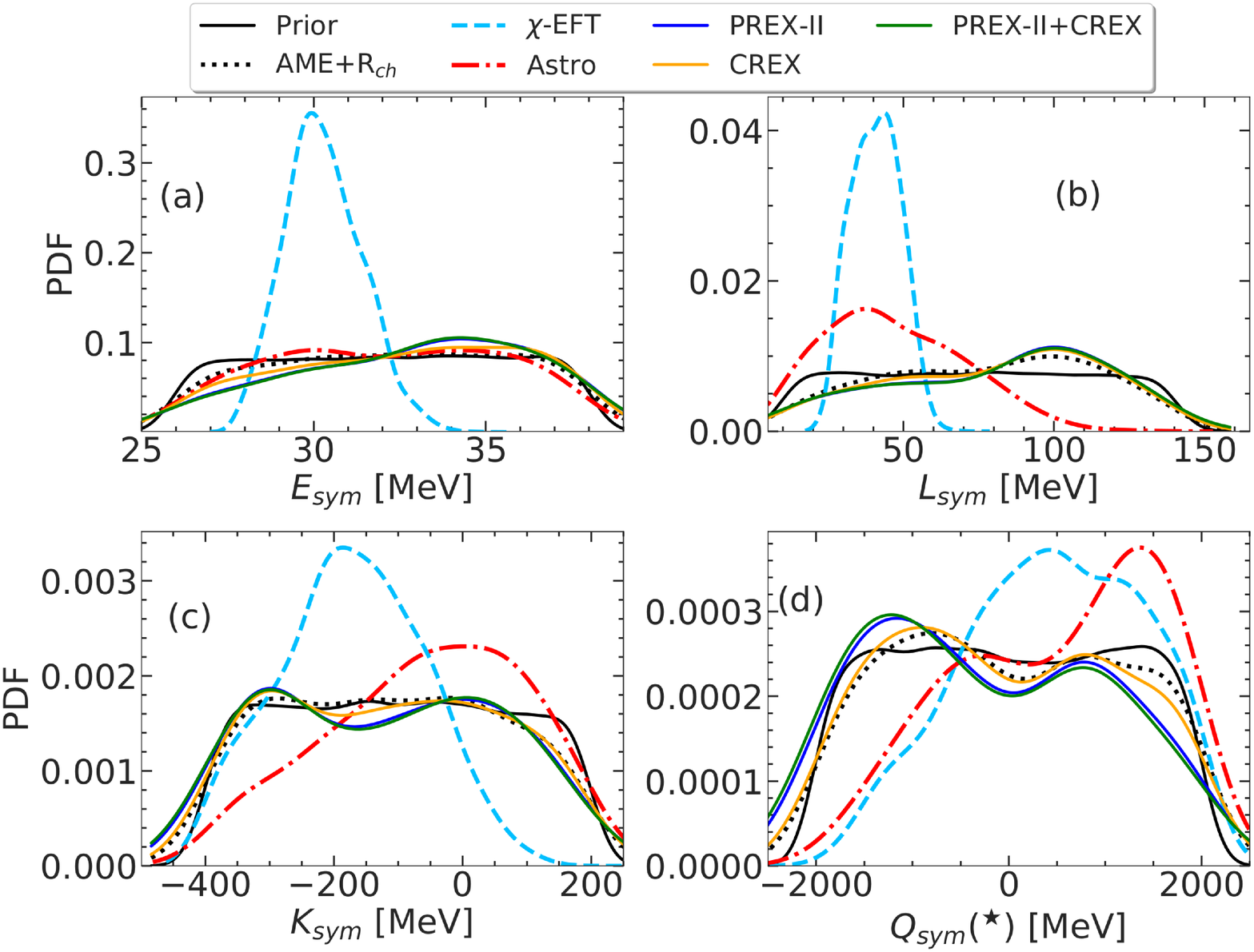}}
\caption{\label{symnm_uncor}
(Color online) Posterior probability distributions of different empirical 
	parameters corresponding to the density behavior of the symmetry energy obtained with 
	different filters described in Section \ref{formalism_filter}. The 
	surface stiffness $Q$ is sampled independently from the bulk parameters in the prior. 
	}
\end{figure}
\begin{figure}[]{}
{\includegraphics[height=2.8in,width=2.8in]{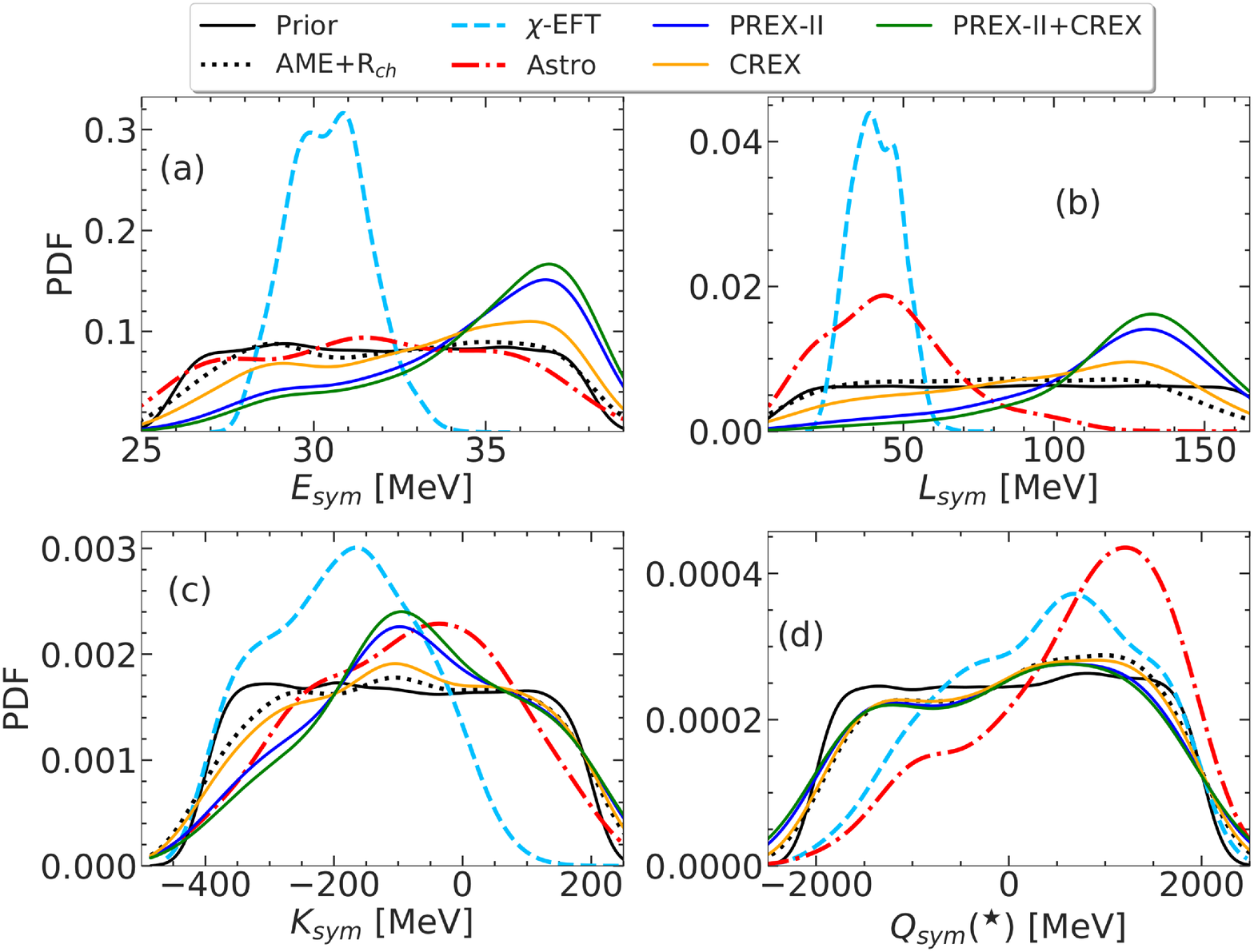}}
\caption{\label{symnm_cor}
(Color online) Posterior probability distributions of different empirical 
	parameters corresponding to the  the density behavior of the symmetry energy, obtained with 
	different filters described in Section \ref{formalism_filter}. The 
	surface stiffness $Q$ prior is obtained following the correlation of Fig. \ref{qlcorr}.
	}
\end{figure}

In Fig. \ref{bech48208} we plot the distribution of binding energies and 
charge radii for $^{208}$Pb  and $^{48}$Ca for different filters using the 
\textit{aETF} method corresponding to the $Q$ sampling of Fig. \ref{qlcorr}. The 
situation is almost identical with the independent sampling of $Q$ (not shown). We depict 
only the cases of $^{208}$Pb and $^{48}$Ca for illustrative purpose and 
mainly because data from PREX-II and CREX correspond to those specific nuclei. The 
corresponding experimental values are indicated by arrows for all four observables. 
Similar behaviors were observed for other nuclei in our nuclear physics informed 
prior \textit{AME+R$_{ch}$}. They are explicitly 
shown in the supplemental material. The reproduction of  ground state radii are not fully satisfactory,
but this limitation is shared by more microscopic mean field studies \cite{Klupfel09, 
Mondal16}. It is known that a precise reproduction of specific charge radii at the mean field level requires 
fine tuning of the interaction \cite{Washiyama12}, {which points towards an important effect 
of beyond mean-field correlations or higher order terms in the functional, 
not linked to EoS properties  \cite{Bender06}. } 
The different filters play almost no role in the distribution the 
binding energies of $^{208}$Pb and $^{48}$Ca (see also supplemental material). 
This is consistent with the observation of Fig. \ref{satnm}(a) and \ref{satnm}(b).
\begin{figure}[]{}
{\includegraphics[height=2.8in,width=2.8in]{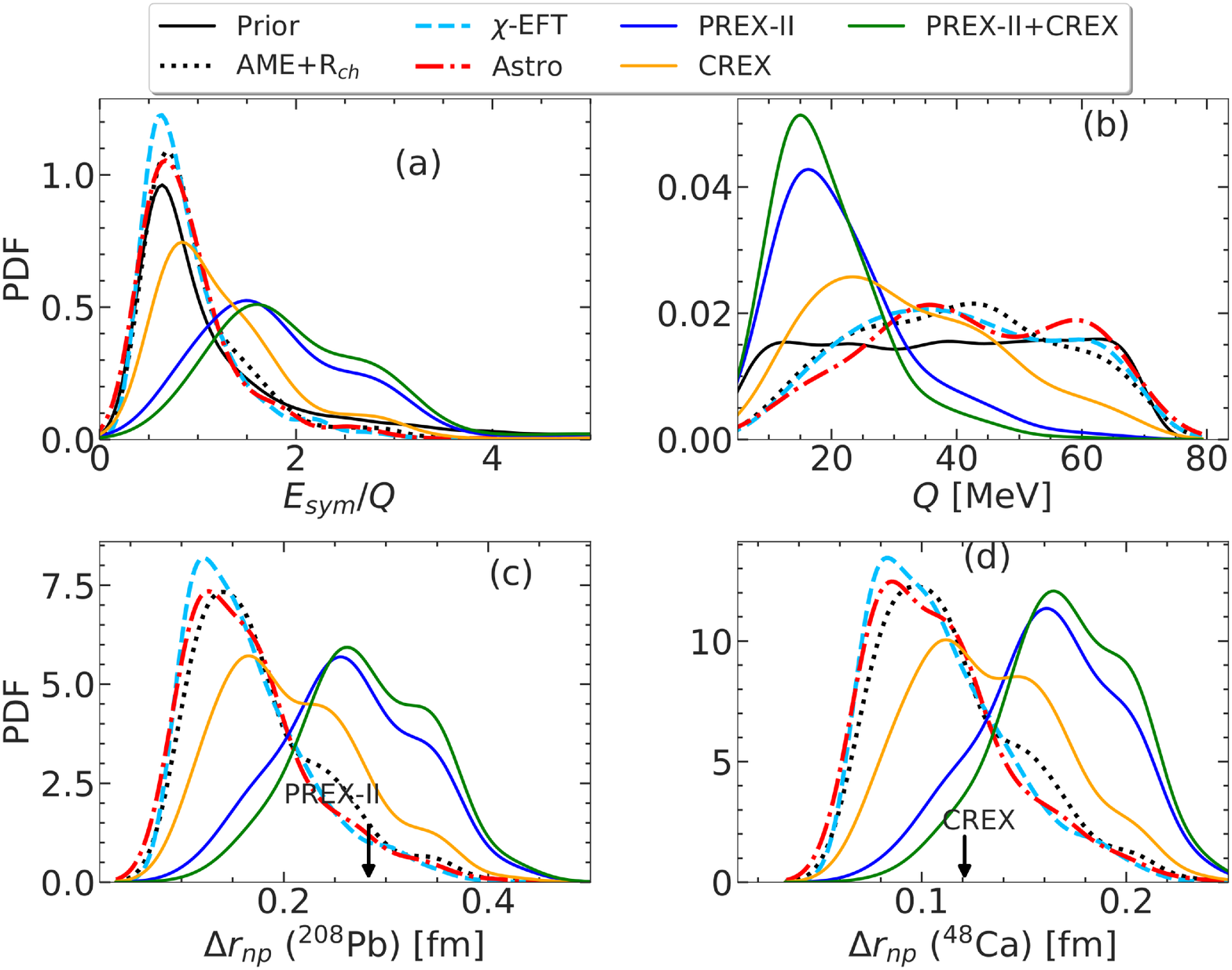}}
\caption{\label{skinlqagnos}
	(Color online) Posterior probability distribution of $E_{sym}/Q$, $L_{sym}$ and 
	$\Delta r_{np}$ of $^{208}$Pb and $^{48}$Ca nuclei obtained with different filters 
	using the models where the surface stiffness $Q$ is sampled independently 
	from the bulk parameters in the prior. 
        }
\end{figure}

Quite distinct behaviors are observed for the symmetry energy parameters \textit{e.g.} 
$E_{sym}, L_{sym}, K_{sym}, Q_{sym}$ and $Q_{sym}^{\star}$ for both the sampling 
methods of $Q$, as shown in Figs. \ref{symnm_uncor} and \ref{symnm_cor}.   
As already reported in previous studies \cite{Dinh-Thi21}, the {$\chi$-EFT} 
calculations offer a fairly precise knowledge on the low order empirical parameters 
in the symmetry energy sector, while this constraint gets relaxed as higher order 
parameters are put to test. Since the \textit{$\chi$-EFT} filter only concerns bulk 
matter properties, the corresponding posteriors are independent of the distribution 
of the surface stiffness parameter $Q$. The impact of astrophysical measurement 
through the \textit{Astro} constraints is also consistent with the one reported in 
the literature \cite{Dinh-Thi21}. 
Models in the prior with very high values of $L_{sym}$ get ruled out because of 
non-existence of meaningful solution of the AME2016 mass table, and the effect is 
particularly important in the correlated sampling technique, where  $L_{sym}$ 
directly impacts the surface properties. In the uncorrelated sample, we observe 
that the constraint from the skin measurements does not affect the values of 
the bulk parameters. Conversely, all symmetry parameters up to order 2 are found 
to be peaked at higher values for the \textit{PREX-II} and \textit{PREX-II+CREX} 
posteriors in the correlated sampling compared to the uncorrelated one, and to the 
corresponding nuclear physics informed \textit{AME+R$_{ch}$} posteriors. It is 
also interesting to note that \textit{Astro} posteriors show distinct peaks in 
$L_{sym}, K_{sym}$ and $Q_{sym}^{\star}$, following rather closely the 
concerned \textit{$\chi$-EFT} posteriors in Figs. \ref{symnm_uncor} and 
\ref{symnm_cor}. 

To understand these distinct behaviors, 
 we show  the neutron skin  and $Q$ distributions in Fig. \ref{skinlqagnos} for agnostic sampling of 
$Q$, and in Fig. \ref{skinlqcorr} for the sampling of $Q$ following Fig. \ref{qlcorr}. 
We can see the effect of the dependence of $Q$ on $L_{sym}$ 
in the correlated sampling technique; very small values of $Q$ getting suppressed 
in the \textit{Prior} and \textit{AME+R$_{ch}$} posterior of Fig.\ref{skinlqcorr}.
A particular parameter of interest to look into is the ratio between symmetry 
energy $E_{sym}$ and stiffness parameter $Q$. It  was shown  
 in  semi-infinite matter calculations across many relativistic and non-relativistic 
interactions \cite{Warda09}, that $E_{sym}/Q$ is linearly correlated with  $\Delta 
r_{np}$ almost in a model independent way. In panel (a) of Fig. \ref{skinlqagnos} and 
\ref{skinlqcorr}, the posterior probability distribution for this ratio is 
depicted for different filters.   
The findings of Ref.\cite{Warda09} are nicely confirmed by our study:
the distribution of $\Delta r_{np}$ for $^{208}$Pb 
[cf. Fig \ref{skinlqagnos}(c) and \ref{skinlqcorr}(c)] and $^{48}$Ca [ 
cf. Fig \ref{skinlqagnos}(d) and \ref{skinlqcorr}(d)] follow very sharply the 
corresponding distribution of $E_{sym}/Q$, respectively.

As in the correlated sampling, small values of $Q$ (and therefore, high values of  
$E_{sym}/Q$ and large skins as measured by PREX-II) are available only for 
large values of $L_{sym}$, the \textit{PREX-II} filter produces 
a stark contrast in the distribution of $L_{sym}$ in Fig.
\ref{symnm_uncor}(b) and \ref{symnm_cor}(b). Specifically, the agnostic sampling of $Q$ makes $L_{sym}$ 
free from skin in Fig. \ref{symnm_uncor}(b), whereas, $L_{sym}$ observes an apex 
in Fig. \ref{symnm_cor}(b) at rather larger values $\sim 120$ MeV with the 
\textit{PREX-II} filter. This value is not far from the value of $L_{sym}=106\pm 37$ MeV, 
inferred by the the authors of Ref.\cite{Reed21} from the PREX data.   It is   important 
to recognize here that, density functionals were employed to infer the $L_{sym}$ 
values in Ref. \cite{Reed21}, and we can therefore expect that a correlation 
similar to the one of Fig. \ref{qlcorr} was present in that study.

\begin{figure}[]{}
{\includegraphics[height=2.8in,width=2.8in]{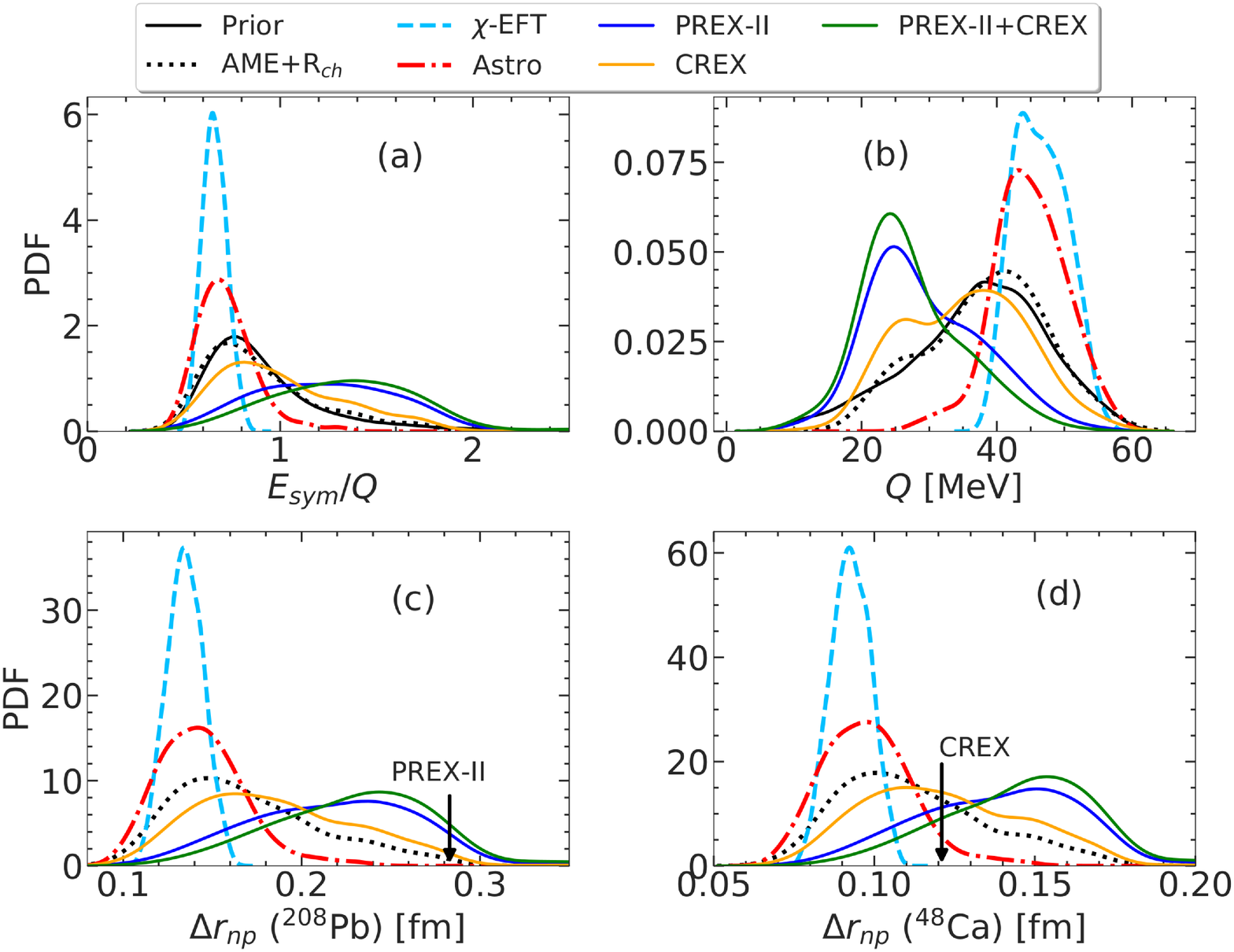}}
\caption{\label{skinlqcorr}
	(Color online) Same as Fig. \ref{skinlqagnos}, but with models where $Q$ and 
	$L_{sym}$ are connected by Fig. \ref{qlcorr}. 
        }
\end{figure}
In the case of the uncorrelated sample in Fig. \ref{skinlqagnos}, the 
\textit{PREX-II} and \textit{CREX} posteriors of $\Delta r_{np}$ are clearly 
peaked on the corresponding experimental values as expected. However, if both 
PREX-II and CREX results are simultaneously taken into account, the prediction for 
the skin of $^{48}Ca$ gets displaced from the experimental value
(see Fig. \ref{skinlqagnos}(d)). This suggests some unintelligible elements 
in the $^{208}Pb$ data, as indicated in the latest paper of the collaboration 
\cite{Adhikari22}. The possible anomaly of the $^{208}Pb$ data is further suggested 
by the results obtained with the correlated sample in Fig.\ref{skinlqcorr}(c). 
Here we can see that imposing the \textit{PREX-II} filter does not result in 
reproducing the PREX-II data satisfactorily, in particular, it fails to reach the higher end 
of the 1-$\sigma$ limit. This means that the $E_{sym}/Q$ distribution is peaked 
on too low values to reproduce the skin measurement. This can be understood 
from the fact that the constraint imposed by the nuclear masses and radii reduce 
the probability of keeping very large values of $L_{sym}$ in the 
correlated sampling, as described in Fig.  \ref{symnm_cor}(b) explicitly. This 
in turn limits retaining very small values of $Q$ (see Fig. \ref{skinlqcorr}(b)) 
and hence large $E_{sym}/Q$ and $\Delta r_{np}$. This result is in qualitative 
agreement with the recent findings of ref. \cite{Reinhard22}, where a difficulty was reported  
in simultaneously reproducing the $^{208}Pb$ and $^{48}Ca$ data with energy 
functionals that give a satisfactory description of masses and charge radii throughout the nuclear chart.

This effect is even more pronounced if we consider the \textit{$\chi$-EFT} filter. 
The theoretical $\chi$-EFT results strongly constrain $L_{sym}$ towards low values, 
and therefore the corresponding posterior is not compatible with the 
PREX-II data, if a strong correlation is assumed between $E_{sym}/Q$ and $L_{sym}$ 
(see Fig.\ref{skinlqcorr}(c),(d)). On the contrary, no strong tension is observed 
if the surface symmetry properties are independently sampled (see Fig. 
\ref{skinlqagnos}(c),(d)). The \textit{CREX} posteriors, on the other hand, have
large overlap with the \textit{$\chi$-EFT} posteriors. 

 Finally, in Fig \ref{skinlqagnos} we can observe that the  astrophysical observations 
(lines noted \textit{Astro}) have negligible impact on the skin prediction following 
closely the \textit{AME+R$_{ch}$} lines, as expected, though it shows some impact on low order 
EoS parameters like $L_{sym}$ and $K_{sym}$ (see Figs. \ref{symnm_uncor}- \ref{symnm_cor}). 
One can observe, however, a strong tension between \textit{Astro} (informed by GW170817 
LIGO-Virgo observation) and \textit{PREX-II} posteriors in Fig. \ref{skinlqcorr}. This 
is primarily manifested through the restriction of smaller $Q$ (Fig. \ref{skinlqcorr}(b)) 
via its \textit{a-priori} assumed correlation with $L_{sym}$ through Fig. \ref{qlcorr}.

\begin{figure}[]{}
{\includegraphics[height=2.8in,width=2.8in]{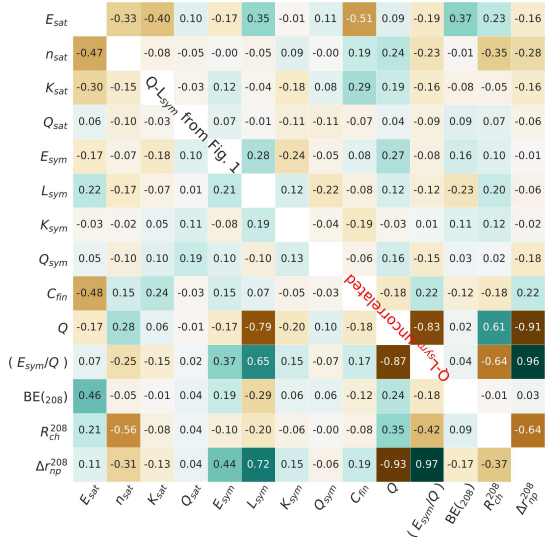}}
\caption{\label{priorcorr}
(Color online) Pearson correlation matrix between various parameters of the 
	metamodel as well as observables of interest obtained with the \textit{AME+R$_{ch}$+PREX-II+CREX} 
	(see text for more details) filter. The numbers below the diagonal correspond 
	to the case where $Q$ and $L_{sym}$ are sampled from Fig. \ref{qlcorr}; 
	the numbers above the diagonal correspond to the independent sampling 
	of $Q$ and $L_{sym}$. 
        }
\end{figure}
The distributions displayed in the previous figures can be interpreted further 
by looking at the correlations among the different observables and parameters.
In Fig. \ref{priorcorr} we show the Pearson correlation coefficients 
between different quantities of interest obtained for the \textit{AME+R$_{ch}$} 
plus \textit{PREX-II+CREX} filter, 
with agnostic sampling of the $Q$ above the diagonal, and with correlated sampling 
of the same below the diagonal. Masses and radii are only correlated to $E_{sat}$ 
and $n_{sat}$, respectively, the latter correlation being highly enhanced 
in the correlated sampling. Absolute correlation between $E_{sym}/Q$ and 
$\Delta r_{np}$ is also noticeable irrespective of the filters and sampling 
techniques \cite{Warda09}. The absence of any other correlations, except a loose 
correlation between $E_{sat}$ and $K_{sat}$, is due to the importance of gradient 
terms that are here treated independently from bulk terms, resulting in a 
significant correlation between $E_{sat}$ and $C_{fin}$ imposed by the mass 
filter. This explains the large posteriors obtained for the different NMP's 
with the \textit{AME+R$_{ch}$} and \textit{Skin}  filters.

The correlations involving $E_{sym}, L_{sym}$ and 
$Q$ in the correlated sampling is a direct consequence of how they were 
populated following the correlation patch of Fig. \ref{qlcorr}.  
It is important to point out here that, we recover the correlation coefficient 
between $Q$ and $L_{sym}$ in the \textit{AME+R$_{ch}$} case ($-0.79$) as we started 
from Fig. \ref{qlcorr} ($-0.77$). 
This clearly explains why a constraint on the skin impacts the properties of the 
symmetry energy, only if the surface stiffness parameter $Q$ is \textit{a-priori} 
well correlated to $L_{sym}$. If this is not the case, as assumed in the uncorrelated 
sample, the filters on the masses, charge radii and skin do not create such correlation, 
and the skin measurement is not constraining for the symmetry energy parameters.

The correlation plot for the \textit{AME+R$_{ch}$} filter alone is very similar 
to the case of \textit{AME+R$_{ch}$+PREX-II+CREX} filter. We display that 
explicitly in the supplemental material. 

\begin{figure}[]{}
{\includegraphics[height=2.8in,width=2.8in]{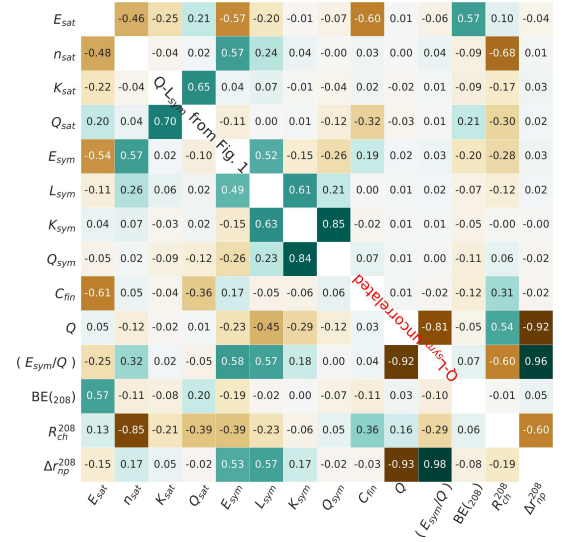}}
\caption{\label{ldcorr}
	(Color online) Same as Fig. \ref{priorcorr}, but obtained with 
	\textit{$\chi$-EFT} filter.
        }
\end{figure}
In Fig. \ref{ldcorr} we display the correlation systematics obtained with 
the \textit{$\chi$-EFT} filter for the same set of observables as in Fig. 
\ref{priorcorr}. Quite recognizably,  the correlation 
imposed \textit{a-priori} between $Q$ and $L_{sym}$ in the correlated sample 
is reduced compared to the \textit{AME+R$_{ch}$} filter. 
Noticeable new correlations among pairs like [$E_{sat}$-$E_{sym}$], 
[$n_{sat}$-$E_{sym}$],  [$E_{sym}$-$L_{sym}$], 
[$L_{sym}$-$K_{sym}$] appear unanimously for the two different sampling 
techniques of $Q$. Similar effect of \textit{$\chi$-EFT} filter on the 
empirical parameters were already observed in \cite{Dinh-Thi21}. Considering 
that in a perfectly known EOS all the coefficients are strongly correlated by 
construction (even if the correlation does not need to be linear), this finding 
measures the amount of information on the behavior of the symmetry energy, 
brought in by the ab-initio calculations of nuclear matter. In particular, 
the strong correlations [$K_{sat}$-$Q_{sat}$] and [$K_{sym}$-$Q_{sym}$] imply 
that, within a Taylor expansion truncated at some low order, the high density 
behavior of the EOS is strongly constrained in the \textit{$\chi$-EFT} posterior, 
even in a density region where the $\chi$-EFT calculations cannot be safely 
extrapolated. This spurious behavior was discussed at length in Ref. 
\cite{Margueron18a}. In that paper, the authors showed that the low density 
behavior is spuriously extrapolated at densities higher than $\approx 0.4$ 
fm$^{-3}$ even if the Taylor expansion is truncated at $N=4$, meaning that 
the low and high density behavior of the EoS are effectively decoupled even 
in a purely nucleonic model. The impact of such spurious extrapolations was 
recently pointed out in ref.\cite{Biswas21}. Conversely, it was shown in  
ref.\cite{Margueron18a} that this problem can be avoided if four extra 
parameters are introduced (see Fig.6 of ref.\cite{Margueron18a}). Those 
parameters, describing the high density behavior and a-priori independent 
of the behavior at saturation, are called $Q_{sat,sym}^\star$ and 
$Z_{sat,sym}^\star$ in this paper.

\begin{figure}[]{}
{\includegraphics[height=2.8in,width=2.8in]{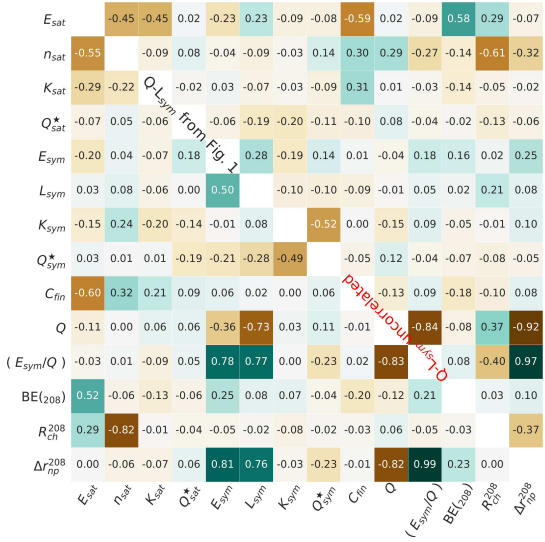}}
\caption{\label{hdlvccorr}
	(Color online) Same as Fig. \ref{priorcorr}, but obtained with 
	\textit{Astro} filter. 
        }
\end{figure}
Correlations obtained with the same set of observables as in 
Fig. \ref{priorcorr}-\ref{ldcorr} obtained with \textit{Astro} filter 
is shown in Fig. \ref{hdlvccorr}. We can see that the correlations between 
the  effective skewness parameters $Q_{sat}^{\star}$, $Q_{sym}^{\star}$ and 
the low order parameters are very different from the ones of Fig. \ref{ldcorr} 
associated to the skewness at and below saturation $Q_{sat}$, $Q_{sym}$, and well 
constrained by nuclear theory. The rest of the correlation plot is fairly 
similar to the one of Fig.\ref{priorcorr}, except for the case between $E_{sym}$ 
and $E_{sym}/Q$. It explains the prominent peaks observed for \textit{Astro} 
posteriors of $E_{sym}/Q$ and $\Delta r_{np}$'s in Fig. \ref{skinlqcorr}. 
But the  overall strong resemblance with \textit{AME+R$_{ch}$} case 
points to the fact that the low density 
physics is strongly decoupled from the high density one. Because of this, 
we expect a limited impact of the skin data on the astrophysical observables. 

\begin{figure}[]{}
	{\includegraphics[height=2.8in,width=2.8in]{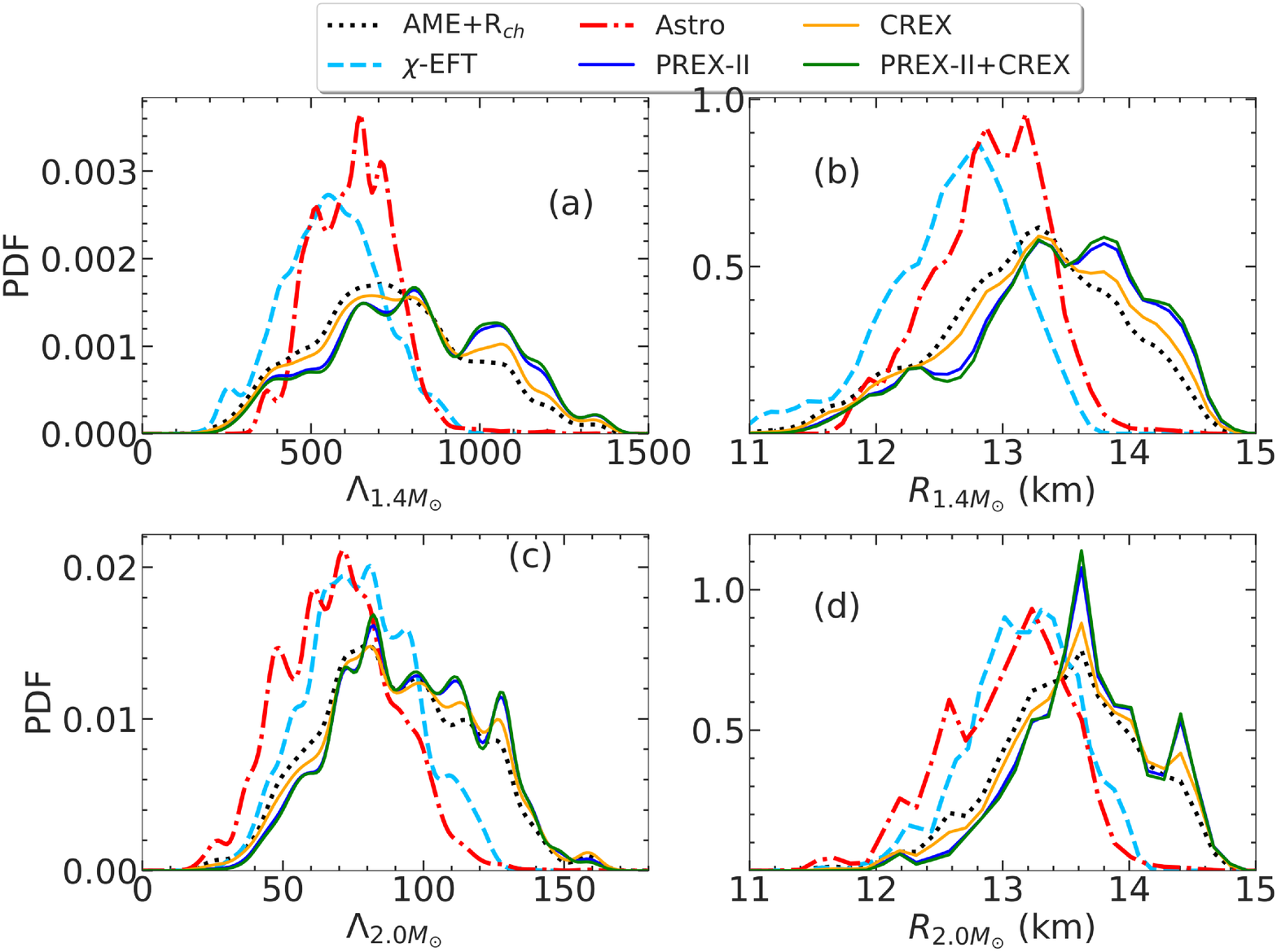}}
\caption{\label{lambdaradiusagnos}
(Color online) Posterior probability distribution of tidal deformability 
	$\Lambda$ (left panels) and radius $R$ (right panels) corresponding 
	to the $1.4M_{\odot}$ (top panels) and $2.0M_{\odot}$ (bottom panels) 
	obtained with the models where surface stiffness $Q$ is sampled independently
       from the bulk parameters in the prior.
        }
\end{figure}
\begin{figure}[]{}
	{\includegraphics[height=2.8in,width=2.8in]{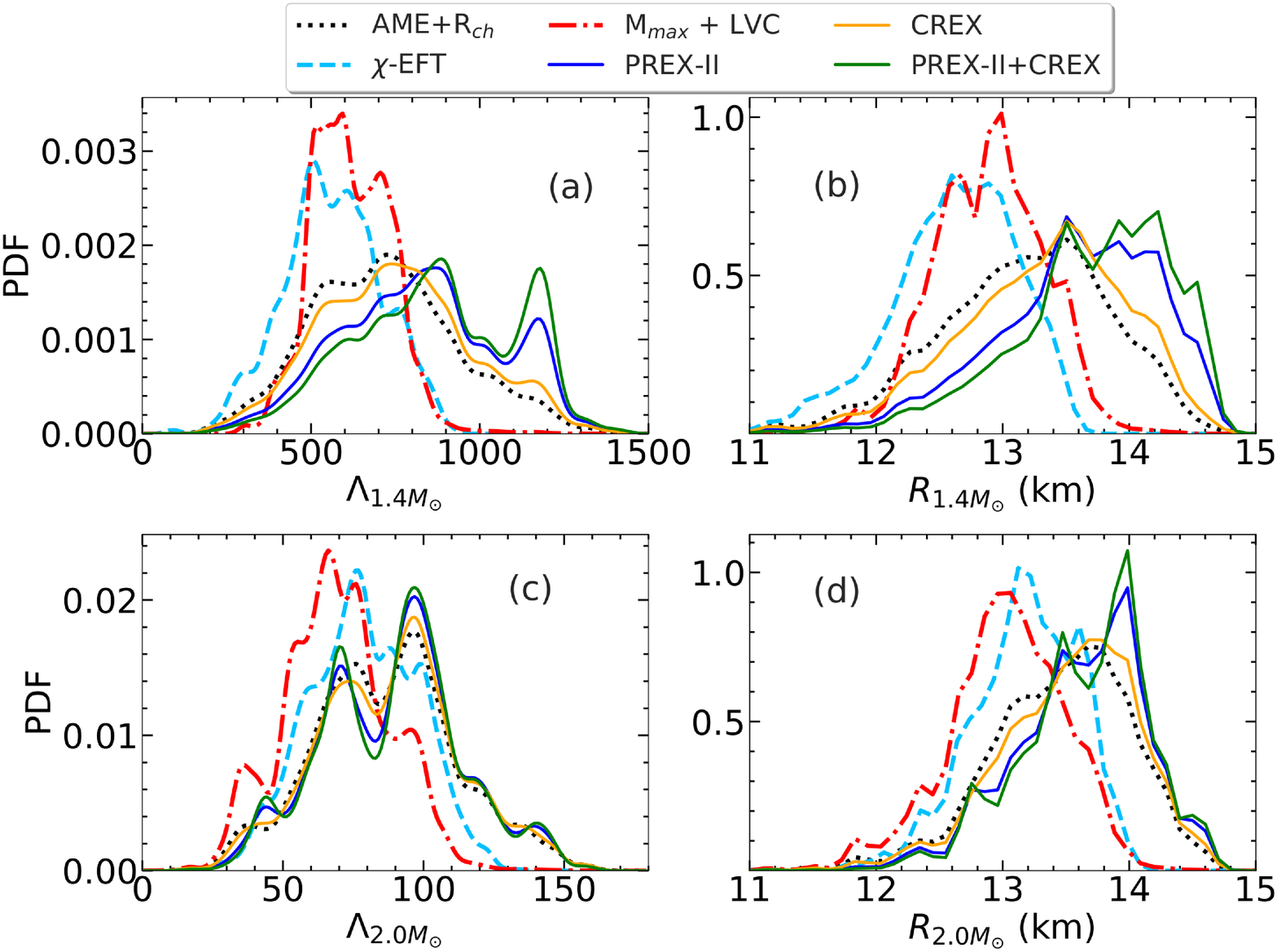}}
\caption{\label{lambdaradiuscorr}
	(Color online) Same as Fig. \ref{lambdaradiusagnos} but with
	$Q$ and $L_{sym}$ correlated through Fig. \ref{qlcorr}.
        }
\end{figure}
This above mentioned point is addressed in Figs. 
\ref{lambdaradiusagnos} and \ref{lambdaradiuscorr}, where we display the 
tidal deformability $\Lambda$ (left) and radii $R$ (right) corresponding 
to 1.4$M_{\odot}$ (up) and 2.0$M_{\odot}$ (down) neutron stars, for the 
agnostic and correlated sampling of $Q$, respectively. As observed before in 
\cite{Dinh-Thi21}, the \textit{$\chi$-EFT} constraint  produces posterior 
distributions of $\Lambda$ and $R$ that are perfectly compatible with the information 
that can be extracted from the astrophysical measurements through the 
\textit{Astro} filter. {This can be interpreted as a demonstration that nucleonic 
degrees of freedom can very well describe the present astrophysical information 
on neutron stars \cite{Dinh-Thi21}, and no evidence of deconfined matter can be 
inferred from the present data on radii and tidal deformability.}

 In all cases, \textit{$\chi$-EFT} and \textit{Astro} filters 
 cut off higher ends of the distributions of the 
experimental nuclear physics informed \textit{AME+R$_{ch}$} distribution.
Coming to the skin measurements, the
\textit{CREX} filter shows absolute insensitivity to the concerned astrophysical 
observables, and the same is true to some extent for \textit{PREX-II} as far 
as the very massive 2.0$M_{\odot}$ neutron star is concerned. This can be easily 
understood from the already discussed effective decoupling between the low and 
high density domain that exists even in the conservative hypothesis of purely 
nucleonic degrees of freedom in the core of neutron stars, as shown in the 
correlation plots above. Concerning the 1.4$M_{\odot}$ neutron star observables,
however, we can observe some effect of the \textit{PREX-II}, and hence 
\textit{PREX-II+CREX} filters,  shifting towards higher values the 
$\Lambda$ and $R$ distributions, particularly in the correlated sampling of $Q$. 
This directs towards an important impact of skin 
measurements on neutron-star observables \cite{Reed21}, and a possible tension
between low density  and high density \textit{$M_{max}$+LVC} data that {was interpreted as pointing } 
towards the existence of a phase transition at high density \cite{Guven20, Biswas21}. 
However, we observe that this tension already appears with respect to the 
\textit{$\chi$-EFT} filter that constrains the same density domain as 
\textit{PREX-II+CREX}, and only appears if the surface parameter $Q$ is robustly 
correlated with $L_{sym}$. Therefore, these findings do not support the interpretation 
of Ref. \cite{Guven20}, and we rather associate this tension to the degree of 
interdependence between bulk and surface properties obtained by the underlying 
nuclear model.  Deeper studies, probably beyond the mean-field picture, are needed 
to sort this problem out comprehensively \cite{Typel10, Typel14, Tanaka21, Reinhard22}.

\section{Conclusions}\label{conclusions}

In this article, we present an upgradation of the nuclear metamodelling technique, 
to calculate the ground state properties of nuclei  within the  ETF method. 
This improvement allows combined and consistent Bayesian analyses of a plethora 
of theoretical and experimental data: not only the constraints from microscopic 
modelling can be treated consistently with the astrophysical observables within 
a unified treatment of the neutron star core and crust, but also nuclear data 
such as binding energies, charge radii and neutron skins can be addressed  on 
the same footing, {though within a simplified semiclassical approach}. 
To this aim, two extra surface parameters are added to the 
parameter space of the bulk metamodelling, namely an isoscalar gradient coupling 
$C_{fin}$, and the surface stiffness parameter $Q$. Within any given homogeneous 
matter functional described through its associated NMP's, finite nuclei density profile are 
described with Fermi functions, with parameters variationally 
obtained within a quasi-analytical version of the ETF theory.

In the present calculation, we particularly concentrated on the consequence of 
CREX and PREX-II results on our understanding of hadronic matter across wide 
range of densities covering both the subsaturation and the supersaturation regimes, 
in the hypothesis that an analytic behavior of the EoS is maintained up to the 
central densities of massive NSs. We addressed the issue of the connection 
between the recent skin data and 
the high density EoS, which is largely debated in the contemporary literature
\cite{Reed21,Reinhard22, Yuksel22, Reinhard21}, in a full Bayesian study
with uncorrelated priors for the  the high order and low order bulk parameters,
and with different hypotheses on the correlations between bulk and surface.

Our main results can be summarized as follows. 
First, comparing the separate constraints coming from both PREX-II and CREX, we  
show that the skin value extracted from PREX-II is hardly compatible with the 
constraints coming from the requirement of reproducing nuclear masses over the 
whole mass range, in qualitative agreement with the conclusions of ref.\cite{Reinhard22}. 
We additionally demonstrate that a possible tension on the preferred values of 
$L_{sym}$ extracted from the observation on tidal deformability from the 
LIGO-Virgo collaboration or the theoretical calculation of low density neutron 
matter using chiral effective field theory, and that from PREX-II, strongly  
depends on degrees of interdependence among bulk (slope $L_{sym}$) and surface 
(stiffness $Q$) parameters of the symmetry energy. To achieve the extremities 
of this interdependence, in one case, we sampled surface stiffness $Q$ and 
$L_{sym}$ independently and in the other, in a correlated manner, as suggested 
by several mean field models \cite{Warda09}. We conclude that the strong 
interplay between bulk and surface symmetry energy parameters 
is the primary reason behind the apparent tension between the 
preferred values of $L_{sym}$ by PREX-II and other experiments or 
observations, while if this correlation is relaxed the tension disappears.

Finally, we critically discuss the impact of observables connected to ground 
state nuclear properties to astrophysical observables that are particularly 
sensitive to densities far beyond the nuclear saturation. We show that the 
subsaturation and supersaturation density domain are effectively decoupled 
even in the simplified nucleonic assumption. This implies that  observations 
from nuclear physics and astrophysics are highly complementary for a full 
understanding of the nuclear equation of state.


\end{document}